\documentstyle[prb,aps,12pt]{revtex}
\begin{document} 
   
\title{Polaron features of the one-dimensional 
Holstein Molecular Crystal Model}
\author{V.~Cataudella, G.~De~Filippis and 
G.~Iadonisi} 
\address{$^\dagger$Dipartimento di Scienze Fisiche, 
Universit\`a di Napoli I-80125 Napoli, Italy}  
\date {\today} 
\maketitle 
\begin {abstract} 
The polaron 
features of the one-dimensional Holstein Molecular Crystal Model 
are investigated by improving a 
variational method introduced recently and based on a 
linear superposition of Bloch 
states that describe 
large and small polaron wave functions. 
The mean number of phonons, the polaron 
kinetic energy, the electron-phonon local correlation function,  
and the ground state spectral weight are calculated and discussed. 
A crossover regime between large and small polaron 
for any value of the adiabatic parameter $\omega_0/t$ is found and a 
polaron phase diagram is proposed.      
\end {abstract} 

\newpage
{\bf Introduction.} In the last decade the idea that the interaction 
of the charge carriers  with the lattice distortions plays an important 
role in determining the electronic and magnetic properties of the 
doped cuprates and manganese oxide perovskites has gained ground more 
and more. Both the infrared spectroscopy and the 
transport measurements, 
involving the colossal magneto-resistance and high $T_c$ 
superconductivity, have pointed out the presence of polaron carriers 
in cuprates and manganites.\cite{1} 

This large amount of experimental data has aroused a renewed interest 
for the Holstein 
molecular crystal model that 
is the simplest model for the study of the interaction 
of a single tight-binding 
electron coupled to an optical local phonon mode.\cite{4} 

For the Holstein Hamiltonian 
the two standard analytical approaches, \cite{5,6} 
based on weak (WCPT) and strong (SCPT) coupling 
perturbation expansions, 
fail to 
describe the region, of greatest physical interest, characterized by 
intermediate couplings and by electronic and phonon energy scales not well 
separated. This regime has been analyzed employing several techniques 
based on Quantum Monte 
Carlo simulations,\cite{7,korn} 
numerical exact diagonalizations of small clusters (EDSC),\cite{8,alexandrov} 
dynamical mean field theory (DMFT),\cite{9} density matrix renormalization 
group\cite{10} and variational approaches.\cite{11,17} 
The general conclusion is that the ground state energy and the effective 
mass in the Holstein model are continuous functions of the electron-phonon 
coupling. 
The ground state properties can change significantly, 
increasing the strength of the interaction,
but without breaking the 
translational symmetry:\cite{lowen}  
there is a smooth crossover between a polaron ground state 
with slightly 
renormalized mass at weak electron-phonon (e-ph) coupling 
and a polaron ground state with a narrow bandwidth at strong 
e-ph coupling. 

In this scenario it has been shown that 
the Global Local variational method\cite{11} (GLVM) 
and the 
Density Matrix Renormalization 
Group\cite{10} (DMRG) approach present some advantages on 
other methods.  
In fact, they provide a very high numerical accuracy when compared, 
for instance, to the Quantum Monte Carlo method (QMCM) that provides 
a numerical accuracy of order of $0.1-0.3\%$. Furthermore the validity 
of these methods is not limited by the use of very small clusters as in the 
numerical exact diagonalization.     
On the other hand we note that: 1) a 
solution of the GLVM for any particular $k$ 
value ($k$ is the wave number of the polaron Bloch state) 
is obtained by minimizing with respect to a very large number 
of parameters, that depends on the number of lattice sites and 
that increases dramatically with increasing the number 
of space dimensions from one to three; 2) the DMRG method involves a 
truncation of the boson Hilbert space and it is based on a heavy numerical 
technique.   

Recently we have proposed a new variational approach\cite{new} 
that is based 
on a linear superposition of Bloch states that describe large and small 
polaron wave functions. 
This approach allows an immediate physical interpretation of the
intermediate regime, does not involve a truncation of the boson
Hilbert space and requires a very little computational effort involving
a very small number of variational parameters that does not depend on the
number of lattice sites and on the dimensionality of the system.

The aim of this paper is to show that a further improvement of 
this method 
allows to give a highly accurate 
description of the polaron 
features for any value of the parameters of the 
Holstein molecular crystal model. In particular, 
the comparison of our results 
with the DMRG and GLVM data points out that the ground state energies 
obtained within our approach are lower than the GLVM energies, 
the difference being about 
0,01\%, and slightly upper than the DMRG energies, 
the difference being about 
0,005\%. 
This agreement strengthens our idea that the true wave function 
is very close to a superposition of the Bloch translationally 
invariant wave functions that provide a very good description of the 
two asymptotic regimes. 
   
{\bf The model}. The Holstein molecular crystal model is described, 
with standard notations, by the 
Hamiltonian:\cite{4} 
\begin{eqnarray} 
&&H=-t\sum_{<i,j>} c^{\dagger}_{i}c_{j}
+\omega_0\sum_{\vec{q}}a^{\dagger}_{\vec{q}}a_{\vec{q}}
+\frac{\omega_0g}{\sqrt{N}}\sum_{i,\vec{q}} 
c^{\dagger}_{i}c_{i}\left[e^{i\vec{q}\cdot \vec{R}_{i}}a_{\vec{q}}
+h.c.\right]~.\nonumber \\
&&
\label{1r}
\end{eqnarray}

As trial wave functions we consider translationally invariant  
Bloch states obtained 
by taking a superposition of 
localized states centered on different lattice sites 
in the same manner in which 
one constructs a Bloch wave function from 
a linear combination of atomic orbitals (see also ref.15): 
\begin{equation}
|\psi^{(i)}_{\vec{k}}\rangle=\frac{1}{\sqrt{N}}\sum_{\vec{R}_n}e^{i\vec{k}\cdot 
\vec{R}_n}|\psi^{(i)}_{\vec{k}}(\vec{R}_n)\rangle
\label{12r}
\end{equation}
where 
\begin{equation}
|\psi^{(i)}_{\vec{k}}(\vec{R}_n)\rangle
=\sum_{\vec{R}_m}c^{\dagger}_{m+n}
\phi^{(i)}_{\vec{k}}(\vec{R}_m)
e^{\sum_{\vec{q}}\left[ f^{(i)}_{\vec{q},\vec{R}_m}(\vec{k})a_{\vec{q}}
e^{i\vec{q}\cdot \vec{R}_n}
-h.c.\right]}|0\rangle
\label{13r}
\end{equation} 
and 
\begin{equation}
f^{(i)}_{\vec{q},\vec{R}_m}(\vec{k})=
f^{(i)}_{\vec{q}}(\vec{k})+\frac{g}{\sqrt{N}}d^{(i)}(\vec{k})
e^{i\vec{q}\cdot \vec{R}_m}~.
\label{14r}
\end{equation}
In the Eq.(\ref{13r}) 
the apex i=l, s indicates the large and small polaron wave function, 
$|0\rangle$ denotes the electronic and boson vacuum state, 
$\phi^{(i)}_{\vec{k}}(\vec{R}_m)$ are variational 
parameters that satisfy the relation: 
\begin{equation}
\sum_{\vec{R}_m}|\phi^{(i)}_{\vec{k}}(\vec{R}_m)|^2=1~,
\label{7r}
\end{equation} 
$d^{(i)}(\vec{k})$ are two variational parameters and 
$f^{(l)}_{\vec{q}}(\vec{k})$ and $f^{(s)}_{\vec{q}}(\vec{k})$ 
represent the phonon distribution functions that are determined 
by minimizing the expectation value of the Hamiltonian (\ref{1r}) on the 
states (\ref{12r}) and by performing, respectively, the limit 
$g\rightarrow 0$ and $g\rightarrow \infty$. 

In this paper we assume: 
\begin{equation}
\phi^{(l)}_{\vec{k}}(\vec{R}_n)
=\alpha^{(l)}_{\vec{k}} \delta_{\vec{R}_n,0}+
\beta^{(l)}_{\vec{k}} \delta_{\vec{R}_n,\vec{\delta}}+
\gamma^{(l)}_{\vec{k}}\delta_{\vec{R}_n,\vec{\eta}}+
\varepsilon^{(l)}_{\vec{k}}\delta_{\vec{R}_n,\vec{\theta}}+
\zeta^{(l)}_{\vec{k}}\delta_{\vec{R}_n,\vec{\vartheta}}
\label{17r}
\end{equation}  
and 
\begin{equation}
\phi^{(s)}_{\vec{k}}(\vec{R}_n)
=\alpha^{(s)}_{\vec{k}} \delta_{\vec{R}_n,0}+
\beta^{(s)}_{\vec{k}} \delta_{\vec{R}_n,\vec{\delta}}+
\gamma^{(s)}_{\vec{k}}\delta_{\vec{R}_n,\vec{\eta}}~.
\label{177r}
\end{equation}  
Here $\beta^{(i)}_{\vec{k}}$, $\gamma^{(i)}_{\vec{k}}$, 
$\varepsilon^{(l)}_{\vec{k}}$ and $\zeta^{(l)}_{\vec{k}}$ 
are variational 
parameters, while 
$\alpha^{(i)}_{\vec{k}}$ are determined in such a way the 
Eq.(\ref{7r}) is satisfied. The symbols   
$\vec{\delta}$, $\vec{\eta}$, $\vec{\theta}$ and $\vec{\vartheta}$ 
indicate, respectively,  
the nearest, the next 
nearest neighbors and so on.  
This choice 
takes into account the broadening of the electronic wave function  
to the nearest neighbors and to the next nearest neighbors 
for the small polaron and to fourth neighbors for the large 
polaron.  

We also note that these wave functions can be systematically 
improved by adding further terms in 
Eq.(\ref{17r}) and Eq.(\ref{177r}). This allows to obtain 
better and better estimates of the polaron energy 
in the two asymptotic limits.    

A careful inspection of these two wave functions shows that, far away from 
the two asymptotic regimes, they are not orthogonal and that the 
off-diagonal matrix elements of the Holstein Hamiltonian are not zero. 
It is then straightforward to determine variationally the polaron ground 
state energy by considering as trial state the linear superposition of the 
large and small wave functions:   
\begin{equation}
|\psi_{\vec{k}}\rangle=\frac{A_{\vec{k}} 
|\overline{\psi}^{(l)}_{\vec{k}}\rangle+
B_{\vec{k}} |\overline{\psi}^{(s)}_{\vec{k}}\rangle}
{\sqrt{A^2_{\vec{k}}+B^2_{\vec{k}}
+2A_{\vec{k}}B_{\vec{k}}S_{\vec{k}}}}
\label{31r}
\end{equation} 
where 
$|\overline{\psi}^{(l)}_{\vec{k}}\rangle$ and 
$|\overline{\psi}^{(s)}_{\vec{k}}\rangle$ are the normalized large and small 
polaron wave functions
and $S_{\vec{k}}$ is the overlap factor.   
In the Eq.(\ref{31r}) $A_{\vec{k}}$ and $B_{\vec{k}}$ 
are two additional variational parameters 
which provide the relative weight of the large 
and small polaron solutions in the ground state of the system for any 
particular value of $\vec{k}$. 
  
The minimization of the quantity 
$\langle\psi_{\vec{k}}|H|\psi_{\vec{k}}\rangle/\langle\psi_{\vec{k}}|\psi_{\vec{k}}\rangle$ 
with respect to the eight variational 
parameters $[\beta^{(s)}_{\vec{k}},\gamma^{(s)}_{\vec{k}},
d^{(s)}(\vec{k})]$ and 
$[\beta^{(l)}_{\vec{k}},\gamma^{(l)}_{\vec{k}},\varepsilon^{(l)}_{\vec{k}},
\zeta^{(l)}_{\vec{k}}, 
d^{(l)}(\vec{k})]$ 
has been performed by making use of a routine 
based on a standard Newton algorithm. 
In this paper 
we will limit our attention to the one-dimensional ground state ($k=0$).  

{\bf Numerical results.} In Fig.1a we plot 
the polaron ground state energy as a function of the e-ph coupling 
constant, in one dimension, for different values of the adiabatic 
parameter $\omega_0/t$. Our variational proposal recovers 
the second order weak and strong 
perturbation results in the two asymptotic 
regimes and, as it is clear from Table 1, provides a highly accurate estimate 
of the polaron ground state energy in the intermediate regime, that is the 
region of greatest physical interest from both experimental and 
theoretical point of view. 

Although the polaron wave function is 
a translationally invariant Bloch state, in the strong coupling limit 
the polaron localization (small polaron) 
appears through a large enhancement of the 
effective mass and through the change in the behavior of the 
correlation 
function between the electron and lattice (see Fig.1c).  
In literature there is unanimous agreement on the conditions leading to 
the small polaron formation. 
It requires $g>1$ and $\lambda>1$, where 
$\lambda$ indicates the ratio between the small polaron binding energy 
and the energy gain of an itinerant electron on a rigid lattice. It is also 
unanimously recognized that decreasing the value of the adiabatic parameter 
$\omega_0/t$ the transition toward the small polaron formation becomes 
sharper and sharper. 

In Fig.1b, 1c, 1d we report the numerical 
results of the mean phonon number, $N$, the e-ph local 
correlation function (the lattice local distortion) scaled with the strong 
coupling result $2g$, $S$,  
and the polaron kinetic energy, 
$K$, in units of the bare electronic kinetic energy, 
as a function of the e-ph coupling constant 
for different values of the adiabatic parameter $\omega_0/t$. 
For weak coupling $K$ is very close to one and $N$ is about 
zero: the electron is slightly affected by the interaction with 
the phonons and drags with itself a phonon cloud that gives rise to a weak 
renormalization of the bare electron mass. $S$ is in agreement with 
the result of the weak coupling perturbation theory and confirms that 
in this regime the extension of the 
polaron is large compared with the lattice parameter of the crystal.   
The non-adiabatic limit deserves to be mentioned, 
where also for weak e-ph coupling 
the extension of the polaron can be less than the lattice parameter 
due to the small value of the  transfer integral $t$.  
Increasing the strength of the 
e-ph interaction the average number of phonons and the lattice local 
distortion increase,   
the kinetic energy reduces and  
asymptotically tends to the values predicted by the strong coupling 
perturbation theory ($N\rightarrow g^2$, $K\rightarrow e^{-g^{2}}$, 
$S\rightarrow 1$). 
In this case the lattice displacement 
is different from zero only on the cell where there is an electron. 
It is worth to point out the behavior of 
the ratio $B/A$ that indicates the relative weight of the 
small and large polaron components in the wave function $|\psi_{\vec{k}}\rangle$: 
by increasing the 
strength of the e-ah interaction it increases and in the crossover 
regime is of the order of the unity indicating that both the wave functions 
contribute to the formation of the so-called intermediate polaron. By 
decreasing the adiabatic parameter $\omega_0/t$, the width of the region 
characterized by values of  
$B/A$ of order of unity decreases confirming that in the 
adiabatic regime the transition toward the small polaron formation becomes 
sharper and sharper as it is also evident from the $K$, $S$ and $N$ 
behaviors. 
      
The conclusions previously drawn and the use of the expressions 
"large, small and intermediate polarons" 
are based, in agreement with previous numerical works,\cite{11} 
on the analysis of 
the polaron ground state properties. 
On the other hand the spectral weight $Z$ associated with the ground state, 
$Z=|\langle\psi_{\vec{k}=0}|c^{\dagger}_{\vec{k}=0}|0\rangle|^{2}$, 
is not entered in the discussion. 
This factor represents the renormalization coefficient of the one-electron 
Green function and gives the fraction of the bare electron state in the 
polaron trial wave function. $Z$ is plotted in Fig.2a. It is evident 
that even in the adiabatic regime the transition toward the small polaron 
formation is accompanied by a very smooth decrease of the ground state 
spectral weight. There is a large region of $g$ values 
where the ground state is well described by an electron with a still 
weakly 
renormalized mass but this state is characterized by a spectral weight 
considerable smaller than the unity: this implies that an essential part 
of the single-particle spectral weight lies at higher energies.  
In particular, in the adiabatic regime although 
the ground state properties, as the kinetic energy and 
the mean number of phonons, change 
sharply in a narrow range of $g$ values, the transfer of spectral 
weight toward the higher energy bands takes place in a smooth and soft way. 
The analysis of the 
spectral weight attached to the polaron lowest energy band allows 
to distinguish 
in the phase diagram three different 
regimes (see Fig.2b): 1) the weak coupling regime or 
the large polaron phase where the ground state is well described by an 
electron with weakly renormalized mass and $Z$ is of order of unity 
($.9<Z \leq 1$). In this limit 
the polaron quasi-particle is well defined and it is 
characterized by coherent motion; 
2) the intermediate polaron phase characterized by 
ground state spectral weight significantly smaller than the unity 
but not negligible ($.1<Z<.9$)
and by renormalized mass larger 
than the bare electronic mass, but not in a dramatic way; 
3) the strong coupling regime or the small polaron phase where the spectral 
weight $Z$ is  
negligible ($0\le Z <.1$); in this limit 
the well known polaron band collapse takes place while 
the polaron extension 
is of order of the lattice constant. Here the main part of spectral weight 
is located at the 
excited states, indicating that the coherent motion is suppressed 
rapidly with increasing the temperature.\cite{emin} 
Consequently we can define two different regimes:        
for electronic and phonon energy scales not well
separated ($\omega_0/t \simeq 1$), the intermediate polaron phase is well 
described by a linear superposition of the two wave functions 
$|\psi^{(l)}_{\vec{k}}\rangle$ and $|\psi^{(s)}_{\vec{k}}\rangle$ and all the ground state 
properties, 
see for instance the kinetic energy behavior in Fig.1d, have intermediate 
values between small and large polaron phases. 
On the other hand, 
in the extreme adiabatic regime the intermediate polaron phase 
is described by the wave function $|\psi^{(l)}_{\vec{k}}\rangle$ but is 
characterized by a small values of the ground state 
spectral weight in the 
single-particle spectrum. In this regime the two ways of characterizing 
the intermediate polaron, based on the analysis of the ground state properties 
and distribution of the spectral weight in the single-particle spectrum, 
are different. 
However, the latter way seems to be more suitable to 
individualize the different polaron regimes in the phase diagram, being 
based on the analysis of a property involving the weight of the 
ground state in the 
single-particle spectrum. 

The existence of a intermediate regime can have interesting consequences on 
the infrared absorption of manganites where it has been shown that the 
conductivity spectra in the metallic region is characterized by the 
presence of a Drude peak at low frequencies and a broad mid-infrared band
around $0.1 eV$.\cite{kim} In fact this features can be explained assuming
(as generally expected\cite{1}) that the absorption is due mainly to polarons
in the intermediate regime. Here since the calculated spectra weight is 
equally distributed between ground and all the excited states, the infrared 
absorption is expected to exhibit two structures with almost equal 
intensity: the coherent( ground-state intraband conductivity) and the 
incoherent (interband conductivity) contributions.        
 
In conclusion, in this paper we have further improved a recently developed 
variational approach to investigate 
the polaron features of the one-dimensional 
Holstein molecular crystal model. 
It has been shown that a simple linear superposition of Bloch states 
that describe the small and large polaron solutions provides an highly 
accurate estimate  
of the ground state energy that is successfully compared with the best 
results available.  It has been also pointed out that a crossover 
regime exists between the large and small polaron phases for any 
value of the adiabatic parameter $\omega_0/t$. It is characterized by a 
soft transfer of spectral weight from the ground state 
toward the higher energy bands.   

\section*{Table 1 captions}
\begin {description} 
\item{Tab.1.} The ground state energies as calculated by various methods 
at $\omega_0/t=1$ and for two different values of the e-ph coupling constant. 
The energies 
(first and second columns) are  
given in units of $\omega_0$.
\end {description}
 
\section*{Figure captions} 
\begin {description} 
\item{Fig.1.} 
The polaron ground state energy (Fig1.a), 
the average number of phonons (Fig.1b), the e-ph local correlation function 
(Fig.1c), the 
polaron kinetic energy (Fig.1d), 
in one dimension,  
are reported as a function of the electron-phonon coupling constant for 
different 
values of the adiabatic parameter $\omega_0 /t$: $\omega_0 /t=2.5$ 
(solid line), $\omega_0 /t=1$ (dashed line), $\omega_0 /t=.5$ 
(dotted line), $\omega_0 /t=.25$ (dashed-dotted line).    
The results obtained within 
the approach discussed in this paper are compared with 
the GLVM data (diamonds), kindly provided 
by A. Romero, 
and the DMRG data (circles),  
kindly 
provided by E. Jeckelmann.  
The energies are given in units of $\omega_0$. The squares indicate 
the values of 
$S$ within the weak coupling perturbation theory. 
\item{Fig.2.} 
In Fig 2.a the ground state spectral weight is reported as a function 
of the e-ph coupling constant 
for the same values of the adiabatic parameter of Fig.1. 
In Fig.2b the polaron 
phase diagram is plotted. The dashed line is characterized by $Z=.9$ and 
separates the large polaron regime from intermediate and small polaron phases. 
The dotted line is characterized by $Z=.1$ and separates the small 
polaron regime from the large and intermediate phases. The dashed-dotted 
line indicates the points in the phase diagram with $Z=.5$.
 
\end {description}

\newpage

\begin{center}
\begin{tabular}{ccc}
\hline \hline
$g=1$         & $g=\sqrt{2}$ & Type of  Method \\
\hline 
-2.000         &  -2.500          & $2^{nd}$ order SCPT (Ref. 4) \\
\hline
-2.4           &  -2.89           & DMFT (Ref. 9)  \\ 
\hline
-2.44721       &  -2.89442        & $2^{nd}$ order WCPT (Ref. 3 )  \\ 
\hline
-2.46770       &  -2.98850        & Previous paper (Ref. 14)  \\
\hline
-2.46931       &  -2.99802        & GLVM (Ref. 11) ($N=32$) \\ 
\hline
-2.46962       &  -2.99833        & this paper ($N=32$)  \\ 
\hline
-2.46968       &  -2.99883        & DMRG (Ref. 10) ($N=32$) \\ 
\hline
-2.471         &  -2.999          & QMCM (Ref. 6)  \\  
\hline
-2.47142       &  -3.00027        & this paper ($N=6$)  \\ 
\hline  
-2.471         &  -3.000          & EDSC (Ref. 8) ($N=6$)  \\ 
\hline \hline
\end{tabular}   
\end{center}

\bigskip
\begin{center}
Table 1
\end{center} 
\end {document}
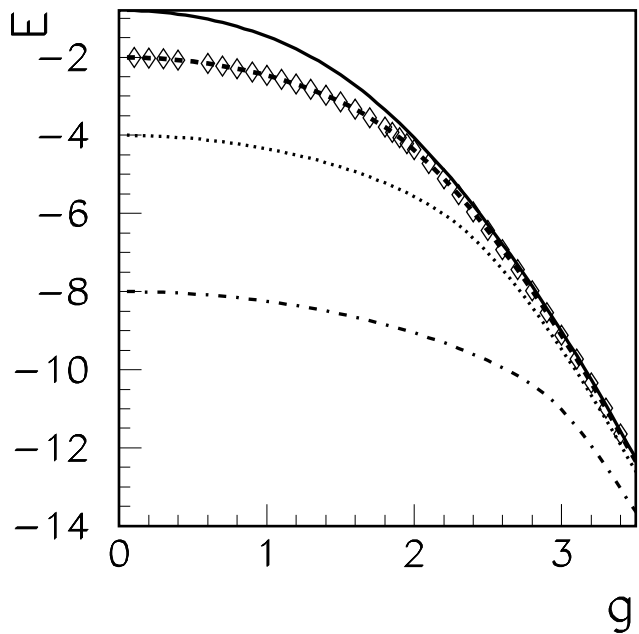
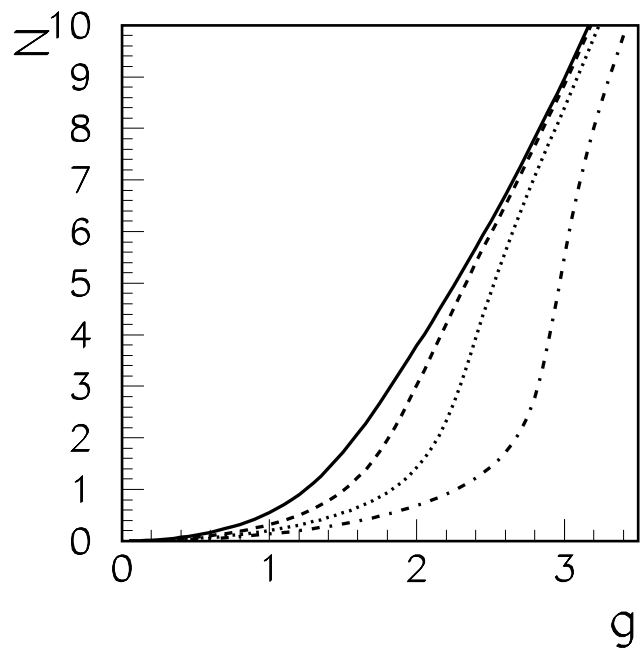
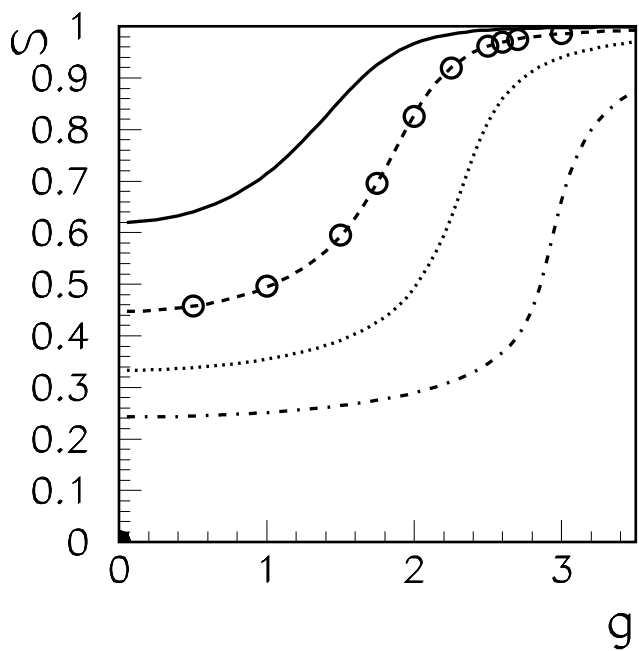
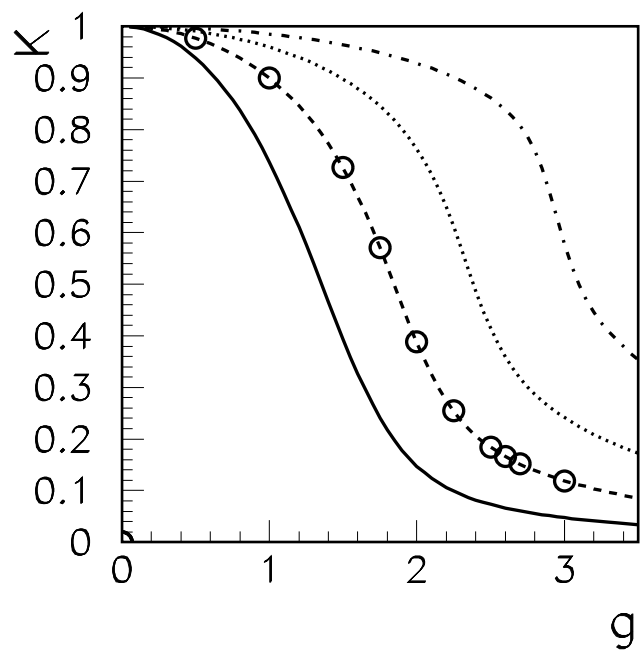
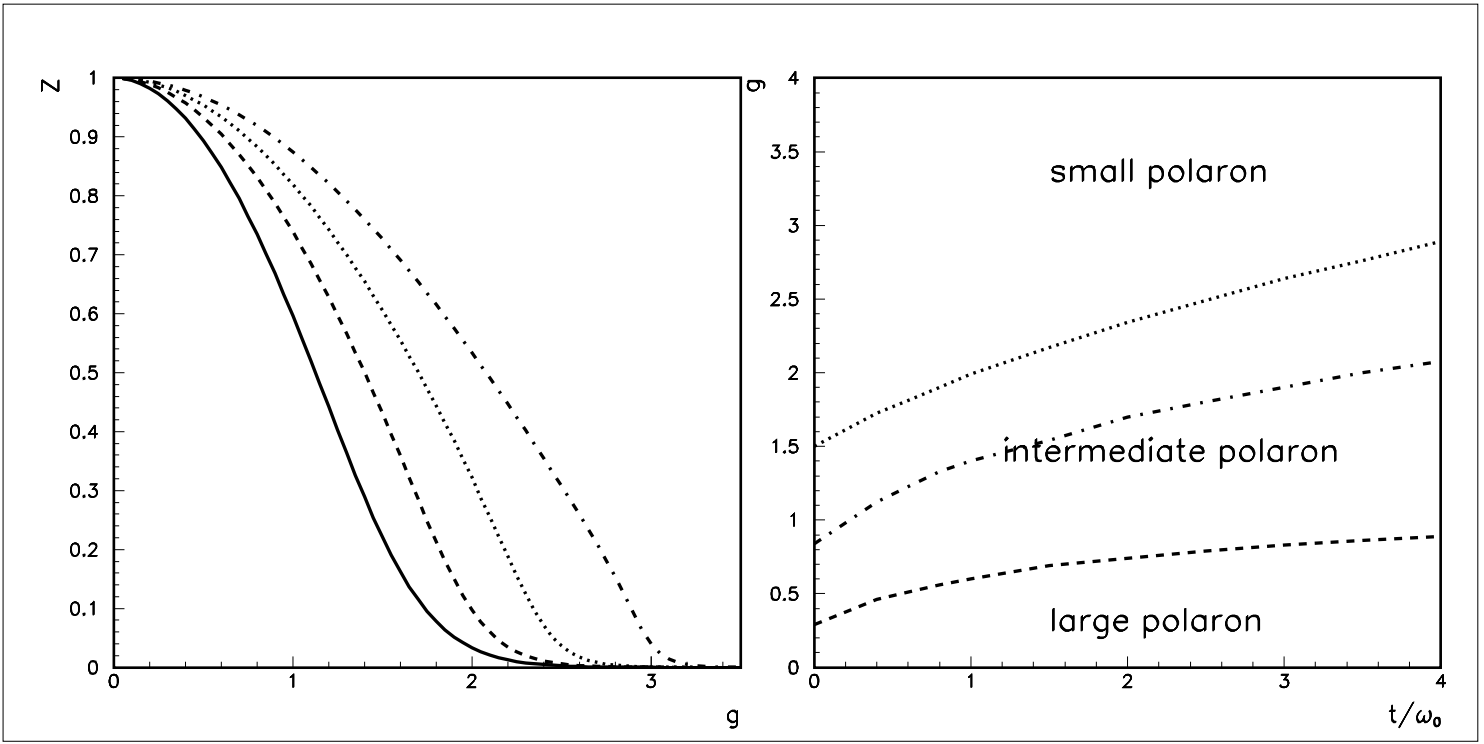
